\documentclass[preprint,12pt]{elsarticle}

\usepackage{amssymb}
\usepackage{amsmath}

\usepackage{multirow}
\usepackage{subfig}
\usepackage{graphicx}
\usepackage{amsmath}
\usepackage{booktabs}
\usepackage{xcolor}
\usepackage{url}

\journal{Nuclear Physics B}

\begin{document}

\begin{frontmatter}



\title{CNN-BiLSTM for sustainable and non-invasive COVID-19 detection via salivary ATR-FTIR spectroscopy}


\author[a]{Anisio P. Santos Junior}
\author[r]{Robinson Sabino-Silva}
\author[r]{Mário Machado Martins}
\author[r]{Thulio Marquez Cunha}
\author[a]{Murillo G. Carneiro}

\affiliation[a]{organization={Faculty of Computing, Federal University of Uberlandia},
            city={Uberlandia},
            postcode={11111}, 
            state={MG},
            country={Brazil}}

\affiliation[r]{organization={Department of Physiology, Institute of Biomedical Sciences, Federal University of Uberlandia},
            addressline={},
            city={Uberlandia},
            postcode={},
            state={MG},
            country={Brazil}}

\begin{abstract}
The COVID-19 pandemic has placed unprecedented strain on healthcare systems and remains a global health concern, especially with the emergence of new variants. Although real-time polymerase chain reaction (RT-PCR) is considered the gold standard for COVID-19 detection, it is expensive, time-consuming, labor-intensive, and sensitive to issues with RNA extraction. In this context, ATR-FTIR spectroscopy analysis of biofluids offers a reagent-free, cost-effective alternative for COVID-19 detection. We propose a novel architecture that combines Convolutional Neural Networks (CNN) with Bidirectional Long Short-Term Memory (BiLSTM) networks, referred to as CNN-BiLSTM, to process spectra generated by ATR-FTIR spectroscopy and diagnose COVID-19 from spectral samples. We compare the performance of this architecture against a standalone CNN and other state-of-the-art machine learning techniques. Experimental results demonstrate that our CNN-BiLSTM model outperforms all other models, achieving an average accuracy and F1-score of 0.80 on a challenging real-world COVID-19 dataset. The addition of the BiLSTM layer to the CNN architecture significantly enhances model performance, making CNN-BiLSTM a more accurate and reliable choice for detecting COVID-19 using ATR-FTIR spectra of non-invasive saliva samples.
\end{abstract}

\begin{graphicalabstract}
    \includegraphics[width=\textwidth]{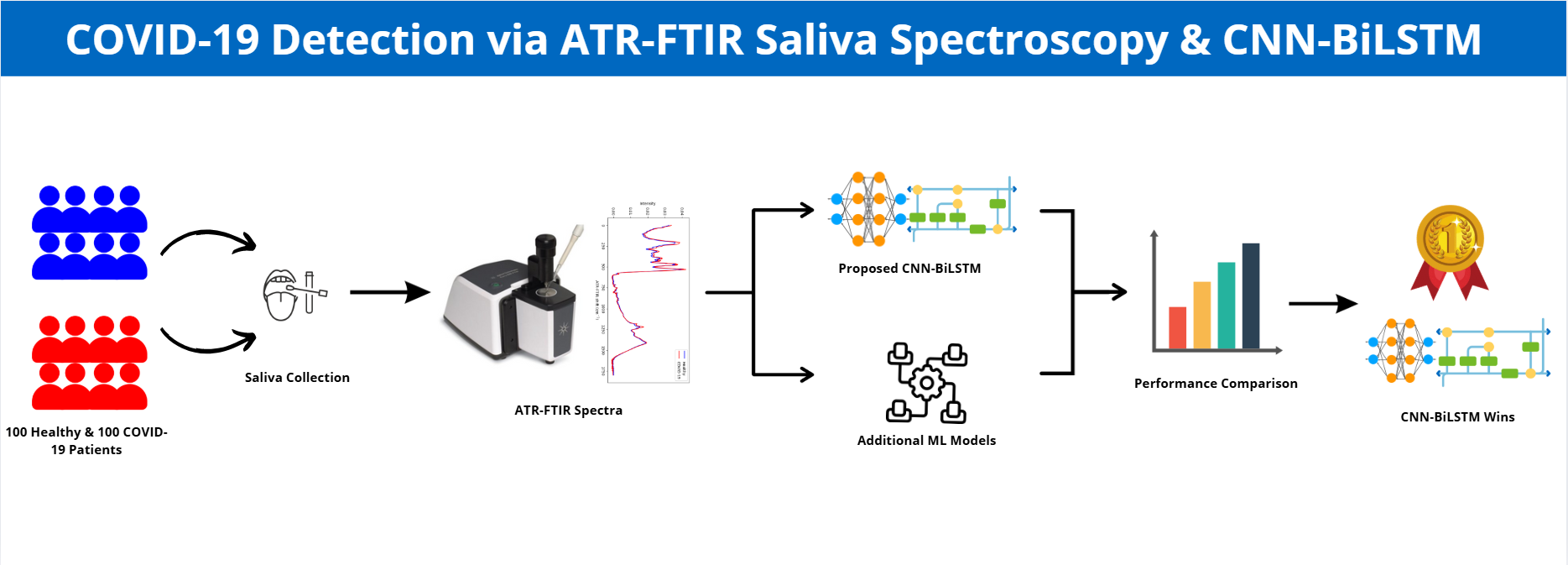}
\end{graphicalabstract}

\begin{highlights}
    \item \textbf{Rapid and Sustainable Detection:} Utilizes ATR-FTIR-based saliva analysis as a reagent-free, fast, and cost-effective alternative for COVID-19 detection compared to RT-PCR.
    
    \item \textbf{Innovative Hybrid Architecture:} Proposes a novel CNN-BiLSTM architecture that combines robust feature extraction from CNNs with the temporal sequence modeling capabilities of BiLSTM.
    
    \item \textbf{Superior Performance:} Experimental results show that the CNN-BiLSTM model outperforms traditional methods—including CNN, NB, RF, XGB, and SVM—achieving an average accuracy and F1-score of 0.80 on a challenging real-world dataset.
    
    \item \textbf{Enhanced Temporal Information Processing:} The addition of the BiLSTM layer significantly improves model performance by effectively capturing temporal dependencies in high-dimensional ATR-FTIR spectral data.
\end{highlights}



\end{frontmatter}



\section{Introduction}
\label{sec1}
The COVID-19 pandemic has highlighted the importance of developing and implementing rapid, accessible, and sustainable testing methods for monitoring infectious diseases. While vaccination is a crucial measure to control the spread of the virus, the inequality in vaccine distribution, especially in underdeveloped countries, and the continuous emergence of virus variants underscore the need for complementary strategies. In this context, mass testing remains an essential tool to mitigate the transmission of COVID-19 \cite{frederick2024}.

To prevent future pandemics and contain potential outbreaks, it is imperative to create new screening techniques that are rapid, cost-effective, and reagent-free. These advancements would be particularly beneficial in environments such as airports, schools, workplaces, and remote communities, where the implementation of effective screening can significantly contribute to public health and safety.

Vibrational spectroscopy, including the ATR-FTIR (Attenuated Total Reflectance Fourier Transform Infrared) technique, is considered a promising approach for biological analyses due to its sustainable, economical, and rapid characteristics. It allows for the extraction of biochemical information without the need for reagents, significantly reducing environmental waste. Furthermore, ATR-FTIR is portable and capable of determining the molecular components of various types of materials quickly and accurately \cite{naseer2021atr,alkhuder2022attenuated}.

Saliva stands out as an interesting biofluid for disease detection because it is sustainable and allows for non-invasive testing. Saliva analysis using ATR-FTIR spectroscopy has been explored in various studies for diagnosing pathological conditions such as Diabetes mellitus \cite{caixeta2023salivary} and Zika virus \cite{oliveira2023salivary}. Vibrational spectroscopy techniques have demonstrated great potential for discriminating between healthy and pathological conditions in biofluids such as blood, urine, saliva, and tissues \cite{naseer2021atr,barauna2021ultrarapid,yin2021efficient}.

The analysis of complex spectral data from ATR-FTIR spectroscopy often benefits from the application of machine learning techniques. These approaches are promising due to their ability to identify subtle patterns in high-dimensional data, which may not be apparent through traditional analytical methods \cite{praja2022attenuated, franca2022atr}. The advantages of using machine learning include the automation of spectra classification, the potential discovery of spectral biomarkers, and improved diagnostic accuracy. Particularly in the analysis of ATR-FTIR data, machine learning has proven effective in distinguishing between healthy and pathological samples, offering a powerful tool for the development of new, rapid, and less invasive diagnostic methods \cite{song2025application, condino2023linear}.

However, many of the machine learning techniques traditionally considered in the literature for ATR-FTIR data analysis, such as Support Vector Machines (SVM), Principal Component Analysis (PCA), and Partial Least Squares Discriminant Analysis (PLS-DA), may present limitations. Although robust, these techniques may not fully capture the complex relationships and sequential characteristics inherent in spectral data. Manual feature engineering can be time-consuming and may not always result in the most informative data representation, especially given the subtlety of spectral variations indicative of pathological states. Furthermore, the generalization ability of these models to larger and more varied datasets can be a challenge \cite{alajaji2025scoping, hasbi2022pattern, wang2019attenuated}. 

To address some of these shortcomings, Convolutional Neural Networks (CNNs) have emerged as a promising alternative for ATR-FTIR data analysis, demonstrating strong performance in learning hierarchical features directly from raw spectral data, thereby mitigating the need for extensive manual feature engineering \cite{jiang2021using, burlacu2021convolutional}. However, while effective at capturing local patterns and hierarchical features, standard CNN architectures may still struggle to explicitly model complex, long-range sequential dependencies across the entire spectrum, which can be crucial for discerning subtle, distributed spectral markers \cite{romero2022towards}.

In this article, we propose a novel architecture for detecting COVID-19 using ATR-FTIR spectra of saliva samples that combines CNN with Bidirectional Long Short-Term Memory (BiLSTM) networks, referred to as CNN-BiLSTM. The hypothesis is that such a combination can potentially improve performance in high-dimensional spectral data classification tasks, given LSTMs' ability to handle temporal sequences and CNNs' robustness in extracting features from complex data. While several works utilize ATR-FTIR spectroscopy with saliva for disease detection, hybrid deep learning architectures are a barely explored topic. The most related works used CNNs for analyzing ATR-FTIR spectra \cite{barauna2021ultrarapid,proquest,ieeexplore}, but without combining them with other deep learning architectures.

In summary, our main contributions are:

\begin{itemize}
    \item Development of a CNN-BiLSTM hybrid architecture for analyzing ATR-FTIR spectra of saliva, incorporating convolutional layers for local feature extraction, bidirectional LSTM units for capturing long-range spectral dependencies, and fully connected layers with dropout regularization for classification.
    \item Comprehensive comparison of predictive performance between the proposed deep learning model and state-of-the-art approaches including Transformer-based architectures, CNN model, traditional machine learning methods, and ensemble techniques on spectral classification tasks.
    \item Systematic evaluation through ablation studies examining the individual contributions of CNN and BiLSTM components, along with controlled experiments analyzing the impact of different hyperparameters, data preprocessing techniques, and the effectiveness of the BiLSTM layer in capturing sequential patterns within high-dimensional ATR-FTIR spectral data.
\end{itemize}

The remainder of the paper is structured as follows: Section \ref{sec:data} presents the description and preparation of the ATR-FTIR spectroscopy dataset, as well as the data preparation and preprocessing steps. Section \ref{sec:model} describes the 1D CNN-BiLSTM proposed in this paper. Section \ref{sec:metric} discusses the experimental setup and evaluation metrics. Section \ref{sec:result} shows the predictive results for COVID-19 detection. Finally, Section \ref{sec:conc} concludes the paper.

\section{Related Works}

In this section, we present the main works related to our research. The application of FTIR spectroscopy for viral detection, particularly for COVID-19, has been a significant area of investigation. Vazquez-Zapien et al. \cite{vazquez2022detection} investigated the use of FTIR spectroscopy of saliva combined with machine learning for COVID-19 detection, analyzing 1275 saliva spectra, including 66 confirmed samples from SARS-CoV-2 virus infected individuals. Various machine learning methodologies were compared, including Multivariate Linear Regression (MLMR), Principal Component Analysis (PCA), Logistic Regression, Support Vector Machine (SVM), and tree-based models. MLMR was identified as the best option for identifying infected individuals, with the shift in the starch I region of the spectrum being crucial for differentiating virus infected patients.

Building on this, Martinez et al. \cite{martinez2021atr} also explored ATR-FTIR spectroscopy of saliva for COVID-19 diagnosis, using 255 samples from COVID-19 patients and 1209 from healthy individuals. Their Multivariate Linear Regression Model (MLRM) highlighted the Amida I (1700--1600~cm$^{-1}$) and IgG (1560--1464~cm$^{-1}$) spectral regions as relevant for discrimination. The study noted increased absorbance in immunoglobulin-related regions and higher DNA/nucleic acid content in the COVID-19 group.

Other bodily fluids and sample types have also been explored. Heino et al. \cite{heino2022diagnostic} assessed the diagnostic potential of ATR-FTIR on 1116 nasopharyngeal swab samples (558 positive, 558 negative for COVID-19), processed by truncating spectra to 1800--900~cm$^{-1}$ and/or 1490--1180~cm$^{-1}$, vector normalization, and averaging. Using Partial Least Squares Discriminant Analysis (PLS-DA) with $k$-fold cross-validation, they reported moderate diagnostic performance, with mean sensitivity of 0.61, and mean specificity of 0.64. The 1490--1180~cm$^{-1}$ spectral region performed similarly to the broader fingerprint region.

Similarly, Barauna et al. \cite{barauna2021ultrarapid} focused on ultrarapid COVID-19 detection using ATR-FTIR analysis of pharyngeal swabs from 111 negative and 70 positive individuals (with a subset used for GA-LDA validation). After preprocessing (truncating to 1800--900~cm$^{-1}$, Savitzky-Golay smoothing, baseline correction, and vector normalization), their Genetic Algorithm-Linear Discriminant Analysis (GA-LDA) model identified spectral features associated with nucleic acids, particularly RNA, as distinctive for COVID-19.

The challenges of data imbalance and feature selection in ATR-FTIR spectral analysis for COVID-19 from saliva were addressed by Hu et al. \cite{hu2024improved}. Using a public dataset of 70 positive and 111 negative saliva samples, they applied baseline correction and Savitzky-Golay smoothing to the 1800--900~cm$^{-1}$ region. They proposed a novel Information Balance (INB) feature selection method alongside data augmentation using WGAN-GP. Comparing classifiers such as KNN, SVM, PLS-DA, and Random Forest, their INB method combined with WGAN-GP achieved accuracies of 88.7\% for the original spectrum and 90.6\% for the second derivative spectrum. However, they noted dataset imbalance and the complexity added by WGAN-GP as limitations.

Investigating the pathophysiological basis, Kazmer et al. \cite{kazmer2022pathophysiological} utilized ATR-FTIR on three biological systems: \textit{in vitro} infected Vero E6 cell supernatants, oral washes from K18-hACE2 inoculated mice, and human saliva from 44 healthy controls and 60 COVID-19 cases. After ethanol treatment and spectral acquisition (4000--650~cm$^{-1}$), baseline adjustment, and normalization, their PLS-DA model predicted COVID-19 in human saliva with 75\% specificity and 93.48\% sensitivity. The study identified significant differences in amide, aliphatic, phosphodiester, and saccharide bands, with a notable increase in the Amide II band indicative of beta-sheet structures.

Beyond COVID-19, FTIR spectroscopy combined with machine learning has shown promise in other diagnostic applications. Lasalvia et al. \cite{lasalvia2023discrimination} combined FTIR spectroscopy with six machine learning algorithms for colorectal cancer diagnosis, using spectra from healthy (FHC) and cancerous (CaCo-2) cells. The models achieved accuracies between 87\% and 100\%, with the neural network outperforming others in accuracy, sensitivity, and specificity.

Wang and Wang \cite{wang2021fourier} summarized several studies that used FTIR spectroscopy for oral cancer diagnosis. Oral cancer is often diagnosed at advanced stages, resulting in high morbidity and mortality. Traditional diagnostic methods are invasive and depend heavily on the operator's skill. FTIR combined with multivariate analysis and machine learning has shown high sensitivity and specificity, although specific numerical values were not provided.

ATR-FTIR analysis of saliva has also been applied to other viral infections in animal models. Guevara et al. \cite{guevara2024salivary} used ATR-FTIR on saliva samples from mice (n=7 vehicle, n=6 CHIKV-infected) to detect Chikungunya virus. Spectra (4000--650~cm$^{-1}$) were preprocessed with baseline correction, Min-Max normalization, truncation to lipid (3050--2800~cm$^{-1}$) and fingerprint (1800--900~cm$^{-1}$) regions, Savitzky-Golay smoothing, and first derivative. Using SVM with five-fold cross-validation, they achieved 83\% accuracy.

Oliveira et al. \cite{oliveira2023salivary} studied Zika virus (ZIKV) using saliva samples processed similarly. Their LDA-SVM model achieved 100\% sensitivity and 75\% specificity (AUC = 0.87), suggesting alterations in lipid, phosphodiester, carbohydrate, and protein components due to ZIKV infection.

Hasbi et al. \cite{hasbi2022pattern} discussed the use of multidimensional spectroscopic data and pattern recognition techniques in UV/Vis, IR, and Raman spectroscopy. Various algorithms, including PCA, Kernel PCA, SPA, GA, PLS-R, LDA, KNN, DT, RF, SVM, PLS-DA, and ANN, were reviewed. The combination of ATR-FTIR with CNN was highlighted for its classification potential, although specific accuracy values were not detailed.

Our work is inspired by several of these studies \cite{shuai2024rapid, vazquez2022detection, martinez2021atr, heino2022diagnostic, barauna2021ultrarapid, hu2024improved, kazmer2022pathophysiological, lasalvia2023discrimination, wang2021fourier, guevara2024salivary, oliveira2023salivary, hasbi2022pattern}, but follows a different approach that considers an architecture composed of CNN and BiLSTM to process ATR-FTIR spectra data. Although previous studies have investigated FTIR spectroscopy with deep learning for disease analysis, our study distinguishes itself by developing one of the first hybrid CNN-BiLSTM approaches for ATR-FTIR. This unique architecture integrates the robust feature extraction ability of CNNs and the BiLSTM’s capacity to represent patterns from sequential data.

\section{Data Description and Preparation}\label{sec:data}

The data used in this work had its collection approved by the local Research Ethical Committee. It is composed of 200 patient samples, with 100 samples from COVID-19 positive patients (COVID-19 group) and 100 samples from individuals with flu-like symptoms (Flu-like group).

Saliva samples were processed using an Attenuated Total Reflectance Fourier Transform Infrared (ATR-FTIR) spectrometer. The ATR-FTIR technique involves the following steps for sample collection and processing:

\begin{itemize}
    \item A small amount of saliva is placed on the ATR crystal, ensuring good contact.
    \item An infrared light beam is directed at the crystal at a certain angle, causing internal reflections. With each reflection, the beam penetrates a short distance into the sample, allowing the measurement of absorbance.
    \item The interaction of the infrared light with the sample generates an absorbance spectrum, which is collected by the detector.
    \item The spectrum is then processed to provide a detailed molecular fingerprint of the sample.
\end{itemize}

\begin{figure}[htbp]
    \subfloat[Average spectrum of the Flu-like group]{\includegraphics[width=0.45\textwidth]{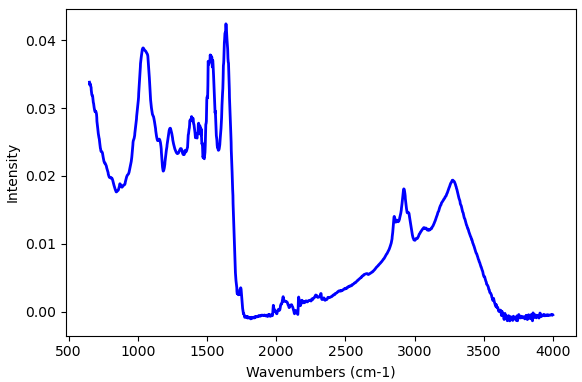}}
    \hspace{10pt}
    \subfloat[Average spectrum of the COVID-19 group]{\includegraphics[width=0.45\textwidth]{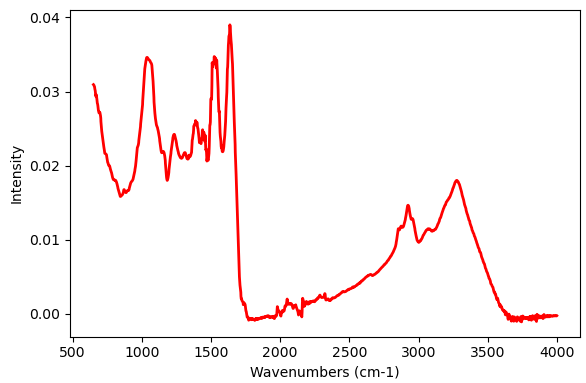}}
    \hspace{10pt}
    \subfloat[Average spectrum of both COVID-19 and Flu-like groups]{\includegraphics[width=0.45\textwidth]{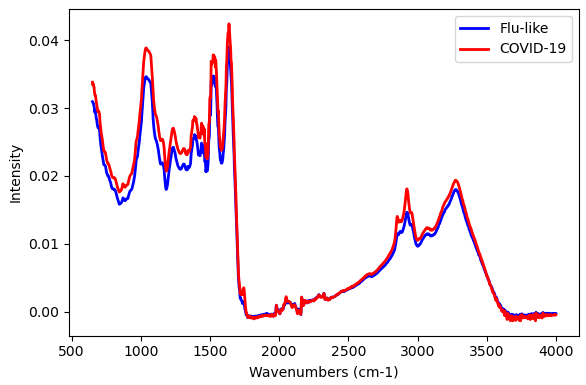}\label{fig:avg}}
    \hspace{10pt}
    \subfloat[Samples from both COVID-19 and Flu-like groups]{\includegraphics[width=0.45\textwidth]{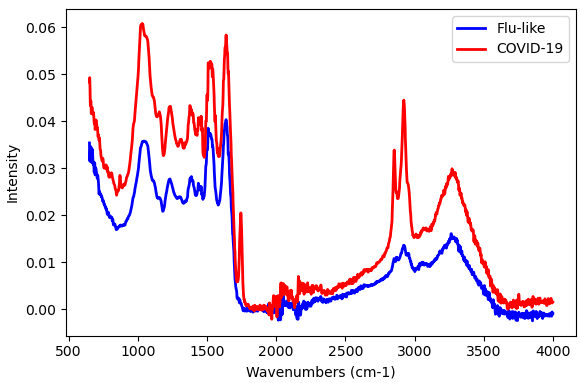}\label{fig:all}}
    \caption{(a) Average spectra of the Flu-like group, (b) Average spectra of the COVID-19 group, (c) Average spectra of COVID-19 and Flu-like groups, (d) All spectra of COVID-19 and Flu-like groups}
    \label{fig:Database_FTIR}
\end{figure}

Figure \ref{fig:Database_FTIR} shows the average spectrum of the Flu-like group (a), the average spectrum of the COVID-19 group (b), the average spectrum of both COVID-19 and Flu-like groups (c), and all spectra of both groups (d). Figures \ref{fig:Database_FTIR}\subref{fig:avg} and \ref{fig:Database_FTIR}\subref{fig:all} highlight distinct regions, showing improved separation between samples from different groups.

Regarding data preparation and preprocessing, two steps were conducted in this study: Savitzky-Golay filtering and vector normalization:
\begin{itemize}
    \item Spectral truncation was performed to focus on the most informative regions of the ATR-FTIR spectrum. Following established practices in spectroscopic analysis, the full spectrum (4000–650 cm$^{-1}$) was truncated to retain only the lipid region (3050–2800 cm$^{-1}$) and fingerprint region (1800–900 cm$^{-1}$), which contain the most relevant biochemical information for viral detection while reducing computational complexity and eliminating less informative spectral regions.
    \item The Savitzky-Golay (SG) filter is used for smoothing and differentiation, optimally fitting a set of data points to a polynomial in the least squares sense \cite{savitzky1964smoothing}. This process is important to reduce noise and smooth the signal while preserving higher-order moments of the original spectrum. SG has been used to normalize and preprocess spectral data from FTIR, Raman, and other spectroscopy equipment due to the inherent presence of noise during sample collection and handling.
    \item Vector normalization is the process where the intensity values of all spectra are normalized by the Euclidean norm.
\end{itemize}

\section{Model Description}\label{sec:model}

This section describes the one-dimensional CNN-BiLSTM proposed in this study for COVID-19 detection using ATR-FTIR spectra from saliva samples. 
The design of a CNN-BiLSTM architecture was motivated by the need to capture both local features and temporal dependencies in ATR-FTIR spectra. Convolutional layers are effective in extracting local features and identifying relevant patterns in the spectrum, while BiLSTM layers are suitable for modeling sequential and temporal dependencies that may exist in the spectral data.

In this study, we hypothesized that BiLSTM is a fundamental component to improve the model's performance. Its ability to process information in both forward and backward directions can allow a more comprehensive understanding of the temporal patterns in the spectra data.

\begin{figure}
    \includegraphics[width=\linewidth]{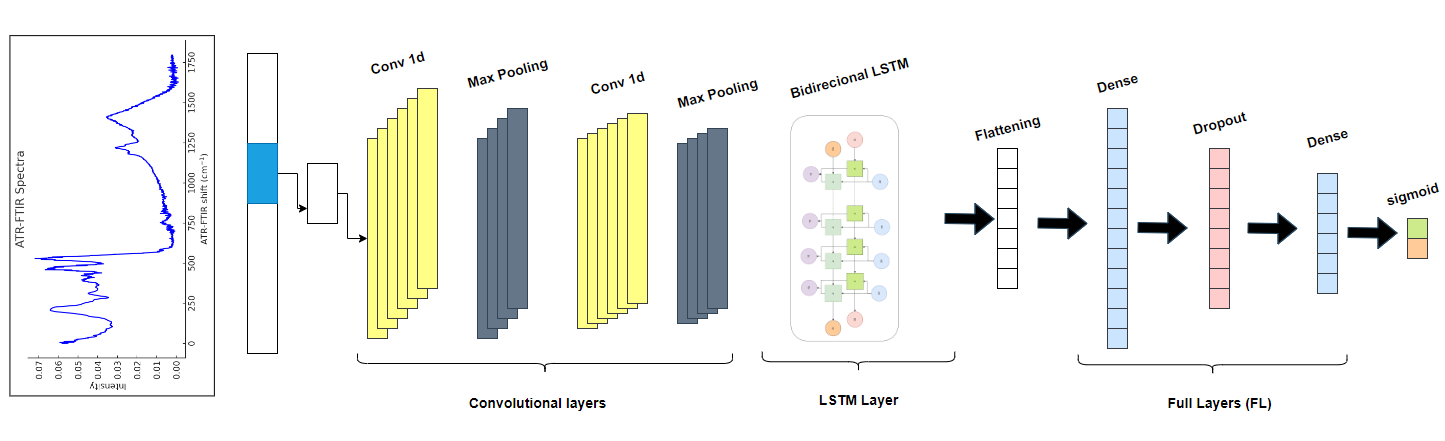}
    \caption{The architecture of the one-dimensional CNN-BiLSTM proposed for COVID-19 detection from salivary ATR-FTIR spectra.}
    \label{CNN_1}
\end{figure}

Figure \ref{CNN_1} presents the layers of the proposed architecture, which, in a few words, is composed of two convolutional layers, a bidirectional LSTM layer and two dense layers. The CNN-BiLSTM architecture receives an ATR-FTIR spectrum as input. Initially, the spectrum passes through two one-dimensional convolutional layers, each one applying a ReLU activation function after the convolution. Between the convolutional layers, MaxPooling layers are used to reduce dimensionality and extract relevant features. After feature extraction, the output of the convolutional layers feeds into a Bidirectional LSTM layer, which sequentially processes the data to capture bidirectional temporal dependencies. The output is then flattened to be used by dense layers that learn the classification with Dropout functions for regularization. The Sigmoid activation function in the final dense layer generates the final prediction.

\subsection{CNN-BiLSTM Architecture Description}
The proposed architecture is detailed in Table \ref{tab:model_architecture}. There are three main types of layers: convolutional, pooling, and fully connected. Additionally, the architecture also includes a bidirectional LSTM layer to capture temporal dependencies in the spectral data. Each of these layers is briefly described next.

\begin{table}[h]
    \centering
    \resizebox{\textwidth}{!}{%
        \begin{tabular}{|c|c|c|}
            \hline
            No. & Layer & Hyper-parameters \\
            \hline
            1 & Conv1D & filters, kernel\_size, activation, kernel\_regularizer = l2(0.001) \\
            2 & MaxPooling1D & pool\_size, strides \\
            3 & Conv1D & filters, kernel\_size, activation, kernel\_regularizer = l2(0.001) \\
            4 & MaxPooling1D & pool\_size, strides \\
            5 & Bidirectional LSTM & units, return\_sequences = True, kernel\_regularizer = l2(0.001) \\
            6 & Flatten & - \\
            7 & Dense & dense\_units, activation, kernel\_regularizer = l2(0.001) \\
            8 & Dropout & dropout\_rate \\
            9 & Dense & dense\_units, activation \\
            10 & Dense & dense\_units = 1, optimizer = sigmoid \\
            \hline
        \end{tabular}%
    }
    \caption{Architecture of the CNN-BiLSTM Model}
    \label{tab:model_architecture}
\end{table}

\subsubsection{Convolutional Layers (Conv1D)} 

Conv1D are designed to learn representative features from the input data. They use filters, also known as kernels, to compute different feature maps. Each neuron in a feature map is connected to a region of neighboring neurons in the previous layer, known as the receptive field \cite{purwono2022understanding}. The feature value at a specific position is calculated as the weighted sum of the inputs in that region, added to a bias term and passed through an activation function:

\begin{equation}
z^l_{i} = \sigma \left( w^{l^T}_{k} x^l_{i} + b^l_k \right),
\end{equation}
where \( x^l_{i} \) is the input centered at position \( i \) in layer \( l \), \( w^{l^T}_{k} \) and \( b^l_k \) are the weight vector and bias term of filter \( k \) in layer \( l \), respectively, and \( \sigma \) is the activation function, typically ReLU (Rectified Linear Units):

\begin{equation}
ReLU(x) = \max{(0, x)}.
\end{equation}

In the proposed architecture, we use L2 regularization to avoid overfitting, especially since the dataset from the three bases is relatively small \cite{metcalf2019strong}. L2 regularization helps penalize large weights in the vector \( w \), encouraging the model to learn smaller and more distributed weights, which can lead to better generalization. The penalty is applied by adding a term to the loss function, which is proportional to the sum of the squared weights:

\begin{equation}
L_{2} = \lambda \sum_{k} w^{2}_{k},
\end{equation}
where \( \lambda \) is the regularization parameter that controls the strength of the penalty. This approach helps prevent the model from overfitting to the training data, promoting better performance on new data.

\subsubsection{Pooling Layer (MaxPooling1D)} 
MaxPooling1D aims to reduce the spatial resolution of the feature maps, achieving shift invariance. This is done by applying a filter that considers only a subset of activations at each stride position, thus reducing the data dimensionality. The most common pooling operation is max pooling, which selects the maximum value in each pooling region, preserving the most salient features \cite{taye2023theoretical}.

\begin{equation}
p_{i,k} = \underset{n=1}{\overset{N}{\max}} \left(z^{l}_{i-n} \right)~,
\end{equation}
where $N$ is the pool size and $z^l_{i}$ is the feature value at position $i$ in the feature map of layer \textit{l}.

\subsubsection{Bidirectional LSTM Layer (BiLSTM)}

BiLSTM is used to capture temporal dependencies in the data by processing the information in both the forward and backward directions. This layer is especially useful for sequential data, such as time series, as it considers the complete context \cite{fan2024bilstm}. Different from time series in which the sequences vary as a function of time, our spectra denote data sequences that vary as a function of infrared wavenumbers. In our architecture, the BiLSTM layer is configured to return full sequences, allowing subsequent layers to utilize all available temporal information. BiLSTM can be defined as follows:

\begin{equation}
h_t = \sigma \left( W_{ih} x_t + b_{ih} + W_{hh} h_{t-1} + b_{hh} \right)~,
\end{equation}
where $h_t$ is the hidden state at position $t$, $W_{ih}$ and $W_{hh}$ are the input and recurrent weights, respectively, $b_{ih}$ and $b_{hh}$ are the bias terms, and $\sigma$ is the activation function.

\subsubsection{Flatten Layer (Flatten)} 

Flatten transforms multidimensional data into a single dimension, preparing it to be passed to the fully connected layers. This operation is necessary to connect the convolutional and pooling layers with the dense layers, which require one-dimensional inputs \cite{educative2024}.

\subsubsection{Fully Connected Layers (Dense)} 

Dense are used to perform the final classification. Each neuron in this layer is connected to all neurons in the previous layer \cite{mocanu2018scalable}. These layers identify the most relevant combinations of features extracted by the previous layers and are defined by:

\begin{equation}
y = \sigma \left( W x + b \right)~,
\end{equation}
where $y$ is the output, $W$ are the weights, $x$ is the input, and $b$ is the bias term. We apply L2 regularization and dropout to avoid overfitting.

\subsubsection{Dropout Layer (Dropout)} 

Dropout is a regularization technique where a fraction of the neurons is randomly deactivated during training \cite{srivastava2014dropout}. This helps prevent overfitting by forcing the network not to rely on specific neurons.

\begin{equation}
y = \sigma \left( W \left( D \odot x \right) + b \right)~,
\end{equation}
where $D$ is a binary mask with the same dimension as $x$, with values 0 or 1 indicating whether a neuron should be deactivated or not.

\subsection{CNN-BiLSTM Hyper-parameters}\label{sec:hyperparameters}
The hyper-parameters used in the CNN-BiLSTM architecture include the number of filters, kernel size, LSTM units, dense units, dropout rate, among others. These hyper-parameters play a crucial role in the architecture and were optimized to maximize the model's performance in the task of ATR-FTIR spectrum classification for COVID-19 detection.

Hyper-parameter tuning was performed using Optuna, a hyper-parameter optimization library based on Bayesian optimization \cite{akiba2019optuna}. Optuna efficiently explores the hyper-parameter search space using techniques such as decision trees to select the most promising combinations. This iterative process allows for the automatic identification of optimal hyper-parameter values, gradually improving model performance over iterations \cite{akiba2019optuna}. Table \ref{tab:hyperparameters} presents the optimized hyper-parameters and their respective value ranges.

\begin{table}[h]
    \centering
    \caption{Hyperparameter Configuration: Values in \{\} are specific choices and values in [] are real values chosen from the specified range.}
    \label{tab:hyperparameters}
    \begin{tabular}{|c|c|}
        \hline
        Hyper-parameter & Values \\
        \hline
        activation & \{relu, tanh\} \\
        dropout\_rate & [0.0, 0.3] \\
        lstm\_units & \{0, 16, 32, 64, 128\} \\
        dense\_units & \{32, 64, 128, 256\} \\
        filters & \{8, 32, 64, 128, 256\} \\
        kernel\_size & \{2, 3, 4, 5\} \\
        pool\_size & \{2, 3, 4, 5\} \\
        strides & \{1, 2, 3, 4\} \\
        optimizer & \{adam, sgd\} \\
        learning\_rate & [1e-5, 1e-1]  \\
        \hline
    \end{tabular}
\end{table}

\section{Experimental Setup and Evaluation Metrics}\label{sec:metric}

In this section, we describe the evaluation metrics used, the training method applied, and the specifications of the machine employed in the experiments for COVID-19 detection using ATR-FTIR spectra.

\subsection{Definition of Evaluation Metrics Used}
The evaluation metrics used in this study were:

\begin{itemize}
    \item \textbf{Accuracy}: Accuracy is the proportion of correct predictions (both true positives and true negatives) over the total number of predictions. It is a simple metric that provides an overall view of model performance.
    \begin{equation}
    \text{Accuracy} = \frac{TP + TN}{TP + TN + FP + FN}
    \end{equation}
    where:
    \begin{itemize}
        \item \(TP\) (True Positives)
        \item \(TN\) (True Negatives)
        \item \(FP\) (False Positives)
        \item \(FN\) (False Negatives)
    \end{itemize}
    
    \item \textbf{Precision}: Precision is the proportion of true positives over the total number of positive predictions (true positives and false positives). It is an important metric when the cost of a false positive is high.
    \begin{equation}
    \text{Precision} = \frac{TP}{TP + FP}
    \end{equation}
    
    \item \textbf{Recall (or Sensitivity)}: Recall, also known as sensitivity, is the proportion of true positives over the total number of actual positives (true positives and false negatives). It is an important metric when the cost of a false negative is high.
    \begin{equation}
    \text{Recall} = \frac{TP}{TP + FN}
    \end{equation}
    
    \item \textbf{Specificity}: Specificity is the proportion of true negatives over the total number of actual negatives (true negatives and false positives). This metric is useful for evaluating the model's ability to correctly identify negative classes.
    \begin{equation}
    \text{Specificity} = \frac{TN}{TN + FP}
    \end{equation}
    
    \item \textbf{F1 Score}: The F1 Score is the harmonic mean of precision and recall. It is useful for evaluating models on imbalanced datasets as it considers both false positives and false negatives. 
    \begin{equation}
    \text{F1 Score} = 2 \cdot \frac{\text{Precision} \cdot \text{Recall}}{\text{Precision} + \text{Recall}}
    \end{equation}
\end{itemize}

These metrics provide a comprehensive view of model performance, allowing for the assessment of its effectiveness in different aspects, such as the ability to correctly identify positive and negative classes and the balance between precision and recall.

\subsection{Evaluation Methods}
Ten simulations were performed using 10-fold cross-validation. In each simulation, each fold served as the test set (comprising 10\% of the data), while the remaining 90\% was used for training; from this training set, 20\% was reserved for validation. In each simulation, Optuna was executed 200 times. The improvement parameter was accuracy, due to the balanced nature of the dataset. At the end, the average of the 10 simulations for each parameter was calculated for the test set.

The computational simulations were conducted using Python on a laptop with the following specifications: 9th generation Core i7 processor, 32 GB of RAM, and a Geforce GTX 1660Ti GPU.

In this study, we compare our CNN-BiLSTM technique against other classification models, such as CNN, Naive Bayes (NB), Random Forest (RF), Extreme Gradient Boosting (XGB), Support Vector Machine (SVM), and a Transformer-based neural network. All models were optimized using Optuna for hyperparameter tuning, following the same optimization framework employed for the CNN-BiLSTM. The CNN shares identical architecture and hyperparameters with the CNN-BiLSTM, with the only difference being the BiLSTM layer adopted by our proposed method. The Transformer architecture was implemented following the methodology described in \cite{wen2022transformers} and subsequently optimized using Optuna. The traditional machine learning algorithms (NB, RF, XGB, and SVM) were implemented using the Python library scikit-learn, with their hyperparameters systematically optimized through Optuna's framework to ensure fair comparison across all methods. The results of these simulations allow us to compare the effectiveness of different models in detecting COVID-19 using salivary ATR-FTIR spectra under equivalent optimization conditions.

\section{Predictive Results for COVID-19 Detection}\label{sec:result}

Table \ref{tab:resultados} summarizes the overall simulation results for the COVID-19 dataset, registering the average performance metrics over the 10 executions. The results demonstrate that the proposed CNN-BiLSTM architecture is highly effective in detecting COVID-19 using ATR-FTIR spectra of saliva samples, outperforming all other tested models. The CNN-BiLSTM achieved an accuracy of 0.80 and an F1-score of 0.80, excelling in all performance metrics. The relatively low standard deviations indicate the model's robustness and consistency across different test runs. Notably, the CNN-BiLSTM achieved the highest accuracy, precision, specificity and F1-score compared to the other models, suggesting a good ability to generalize to new data.

\begin{table}[h]
    \centering
    \resizebox{\textwidth}{!}{%
        \begin{tabular}{|c|c|c|c|c|c|}
            \toprule
            \textbf{Model} & \textbf{Accuracy} & \textbf{Precision} & \textbf{Sensitivity} & \textbf{Specificity} & \textbf{F1-Score} \\ \midrule
            NB             & 0.50$\pm$0.1066     & 0.50$\pm$0.0757      & 0.72$\pm$0.1619        & 0.27$\pm$0.1636        & 0.58$\pm$0.1619     \\ 
            RF             & 0.55$\pm$0.1012     & 0.56$\pm$0.1173      & 0.54$\pm$0.1713        & 0.55$\pm$0.2321        & 0.54$\pm$0.1713     \\ 
            XGB            & 0.54$\pm$0.1248     & 0.53$\pm$0.1377      & 0.51$\pm$0.2079        & 0.56$\pm$0.1897        & 0.51$\pm$0.2079     \\ 
            MLP            & 0.54$\pm$0.1292     & 0.53$\pm$0.1273      & 0.51$\pm$0.2079        & 0.56$\pm$0.1265        & 0.51$\pm$0.2079     \\ 
            SVM            & 0.54$\pm$0.0937     & 0.52$\pm$0.0642      & \textbf{0.88$\pm$0.1932} & 0.20$\pm$0.1333        & 0.65$\pm$0.1932     \\ 
            CNN            & 0.76$\pm$0.0568     & 0.74$\pm$0.0696      & 0.83$\pm$0.1252        & 0.69$\pm$0.1197        & 0.77$\pm$0.1252     \\ 
            Transformer    & 0.67$\pm$0.1107     & 0.70$\pm$0.1631      & 0.65$\pm$0.1434        & 0.68$\pm$0.2150        & 0.66$\pm$0.1042     \\
            \textbf{CNN-BiLSTM}       & \textbf{0.80$\pm$0.0926} & \textbf{0.80$\pm$0.1254} & 0.82$\pm$0.1033        & \textbf{0.77$\pm$0.1567} & \textbf{0.80$\pm$0.0860} \\ \bottomrule
        \end{tabular}%
    }
    \caption{Averaged results of the proposed CNN-BiLSTM architecture in comparison with CNN,Transformer and other classification algorithms.}
    \label{tab:resultados}
\end{table}

By analyzing the CNN results in Table \ref{tab:resultados}, which achieved an accuracy of 0.76 and an F1-score of 0.77, it is evident that the inclusion of the BiLSTM layer provided a significant improvement in the results. This performance boost highlights the advantage of incorporating temporal information in the modeling, crucial for analyzing ATR-FTIR spectra. Additionally, the lower standard deviations in the CNN-BiLSTM model emphasize its greater reliability and effectiveness.

The Transformer model achieved intermediate results, with an accuracy of 0.67 and F1-score of 0.66. Although the Transformer provided better performance than most traditional machine learning algorithms tested, it was still outperformed by the CNN-based models. The higher standard deviations observed indicate that the Transformer model demonstrated less consistency across different executions, suggesting potential limitations in capturing the specific spectral patterns of ATR-FTIR data compared to CNN-based models.

It is important to note that, although the CNN-BiLSTM model outperformed the others in almost all metrics, the SVM exhibited the highest sensitivity, with 0.88$\pm$0.1932. This indicates that the SVM was particularly effective in identifying positive COVID-19 cases, although it did not perform well in other metrics. In fact, the SVM's low specificity (0.20$\pm$0.1333) implies that nearly 80\% of flu-like individuals would be misclassified as COVID-19 positive, rendering it clinically unsuitable.

\begin{figure}
    \centering
        \includegraphics[width=0.48\textwidth]{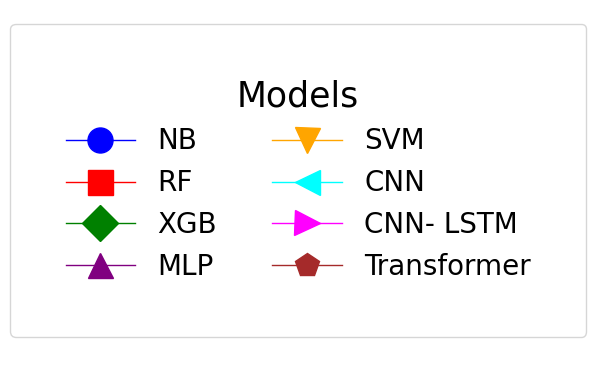}
        \hspace{4pt}
        \subfloat[Accuracy]{\includegraphics[width=0.48\textwidth]{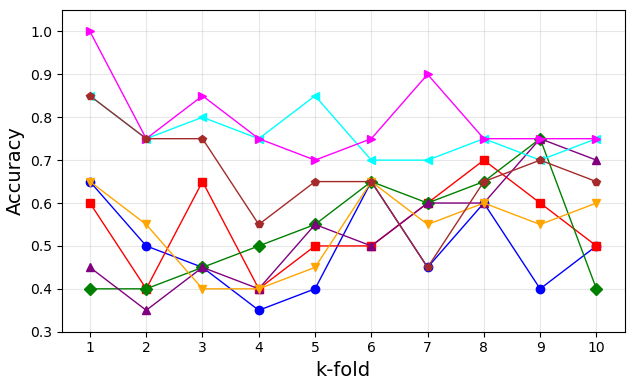}} \\
        \subfloat[Precision]{\includegraphics[width=0.48\textwidth]{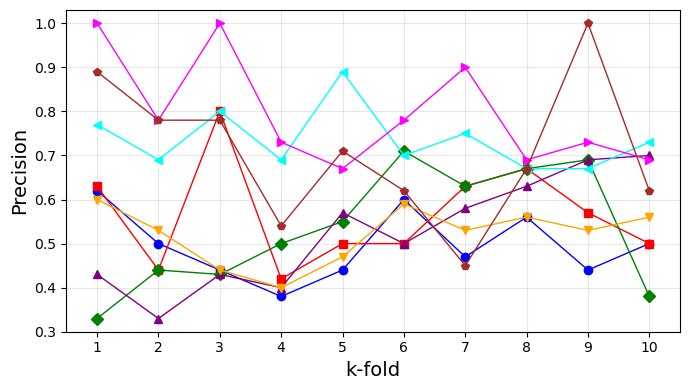}}
        \hspace{4pt}
        \subfloat[Sensitivity]{\includegraphics[width=0.48\textwidth]{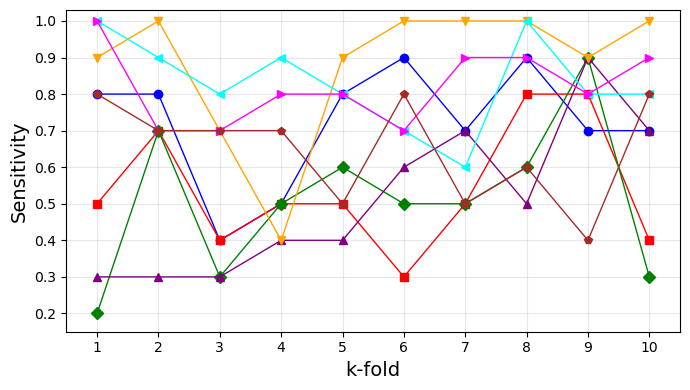}}
        
        \subfloat[Specificity]{\includegraphics[width=0.48\textwidth]{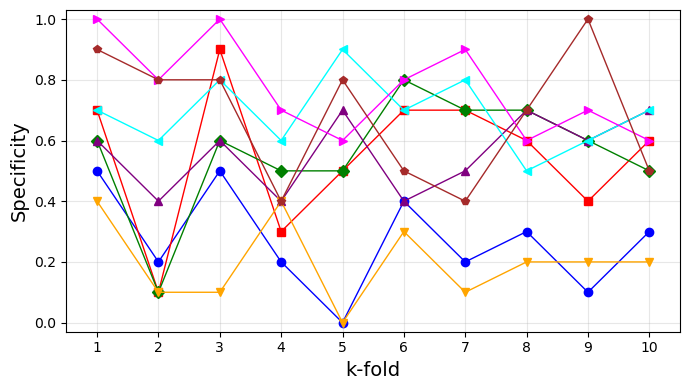}}
        \hspace{4pt}
        \subfloat[F1 Score]{\includegraphics[width=0.48\textwidth]{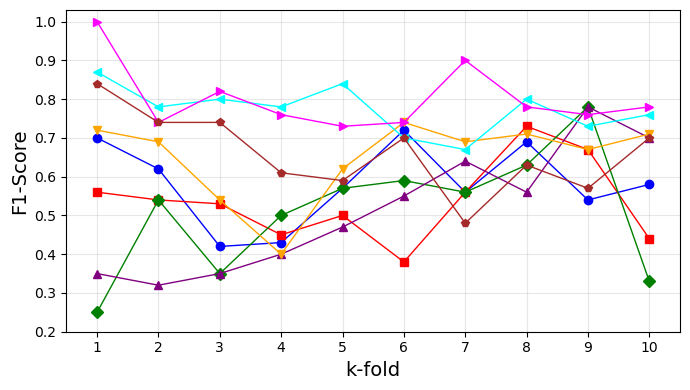}}
    \caption{Analysis of the predictive performance of the techniques under comparison over the 10-fold cross-validation.}
    \label{fig:k_fold_splits_test}
\end{figure}

Figure \ref{fig:k_fold_splits_test} shows the predictive performance of each technique under comparison averaged over the 10 simulations of 10-fold cross-validation considering each one of the metrics. As one can see, CNN-BiLSTM achieved sound results, always among the top predictive performance in each fold considering all the metrics under analysis. For example, the accuracy and F1 score graphs clearly show that the CNN-BiLSTM demonstrated superior performance compared to the other models, maintaining high accuracy and F1 score rates over the folds, while other machine learning models showed greater variation and instability in metrics.

In summary, the CNN-BiLSTM demonstrated to be the best choice for this dataset, combining the deep learning capability of the CNN with the BiLSTM's ability to capture temporal dependencies, resulting in a robust and highly effective model for COVID-19 detection.

\subsection{Visualization of Feature Representation using t-SNE}

To gain further insight into how the CNN-BiLSTM model processes the spectral data, we employed t-Distributed Stochastic Neighbor Embedding (t-SNE) \cite{van2008visualizing} for visualization. t-SNE is a dimensionality reduction technique particularly well-suited for visualizing high-dimensional datasets in a low-dimensional space (typically 2D or 3D), revealing underlying structures and cluster separations.

\begin{figure}
    \includegraphics[width=\linewidth]{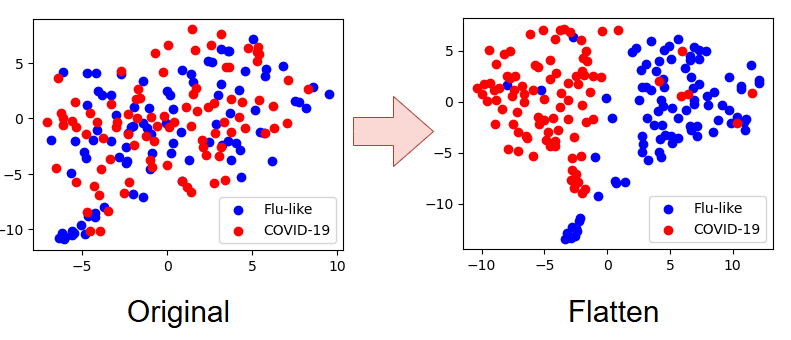}
    \caption{t-SNE visualization comparing the original preprocessed ATR-FTIR spectral data (Left) with the data features extracted by the CNN-BiLSTM model after the Flatten layer (Right). Blue points represent the Flu-like group, and red points represent the COVID-19 group.}
    \label{fig:tsne_original_flatten}
\end{figure}

Figure \ref{fig:tsne_original_flatten} presents the t-SNE visualization comparing the projection of the original preprocessed ATR-FTIR spectra (Left) with the projection of the features extracted by the CNN-BiLSTM model immediately after the Flatten layer (Right), before the final classification layers. Points are colored according to their class: blue for the Healthy/Flu-like group and red for the COVID-19 group.

Observing the left panel of Figure \ref{fig:tsne_original_flatten} ("Original"), the t-SNE projection of the input data shows a significant degree of overlap between the Flu-like (blue) and COVID-19 (red) samples. While some subtle grouping might exist, the classes are heavily intermingled, visually demonstrating the challenge of separating these groups based solely on the original spectral data using simpler linear models.

In stark contrast, the right panel of Figure \ref{fig:tsne_original_flatten} ("Flatten") reveals a dramatically improved separation between the classes after the data has been processed through the convolutional and BiLSTM layers of our proposed model. The COVID-19 samples (red) form a relatively distinct cluster, primarily located in the lower portion of the plot, while the Flu-like samples (blue) are predominantly grouped in the upper portion. Although a few points are still mixed, the transformation learned by the CNN-BiLSTM network has clearly mapped the input data into a new feature space where the two classes are substantially more separable. This visual evidence strongly suggests that the feature extraction part of the model (CNN layers combined with BiLSTM) is effectively learning discriminative patterns within the complex spectral data, which facilitates the subsequent classification task performed by the dense layers. The enhanced clustering seen in the t-SNE plot aligns with and helps explain the superior quantitative performance metrics achieved by the CNN-BiLSTM model, as detailed in Table \ref{tab:resultados}.

\subsection{Explainability analysis}

For a deeper understanding of the model's decision-making process based on the ATR-FTIR spectral data, SHapley Additive exPlanations (SHAP) \cite{NIPS2017_7062} analysis was employed. SHAP provides a unified framework to explain the output of any machine learning model. By calculating the contribution of each input feature (i.e., specific wavenumbers or spectral features) to the prediction for an individual sample, SHAP values reveal which parts of the spectrum are most influential in driving the model's output, thereby enhancing the interpretability of the spectral-based analysis.

\begin{table}[htbp]
\centering
\caption{ATR-FTIR spectral peaks (in cm$^{-1}$) associated with major biomolecular constituents in biological samples. The plot delineates characteristic vibrational modes of lipids, proteins, and nucleic acids.}
\begin{tabular}{|c| p{5.5cm}|l|}
\hline
\textbf{Wavenumber (cm$^{-1}$)} & \textbf{Assignment} & \textbf{Biomolecular} \\
\hline
2922 & Asymmetric CH$_2$ stretching of acyl chains & Lipids \\
2920 & Asymmetric CH$_2$ stretching of acyl chains & Lipids \\
1769 & Unassigned band & --- \\
1744 & Ester group vibration of triglycerides & Lipids \\
1696 & Antiparallel $\beta$-sheet (Amide I) & Proteins \\
1674 & C=O stretching (Amide I) & Proteins \\
1670 & C=O stretching (Amide I) & Proteins \\
1375 & C-N stretching of cytosine & Nucleic acids \\
1374 & C-N stretching of cytosine & Nucleic acids \\
1312 & Amide III band components & Proteins \\
\hline
\end{tabular}
\caption*{ATR-FTIR spectral peaks (in cm$^{-1}$) associated with major biomolecular constituents in biological samples. The plot delineates characteristic vibrational modes of lipids, proteins, and nucleic acids.}
\label{tab:ftir_peaks}
\end{table}

ATR-FTIR spectral analysis identifies specific wavenumber regions that correspond to well-defined vibrational modes of biomolecules found in saliva. As detailed in Table \ref{tab:ftir_peaks}, the bands at 2920 and 2922 cm$^{-1}$, linked to asymmetric CH$_2$ stretching, are characteristic of lipid acyl chains and serve as markers of the salivary lipid profile. The strong absorption at 1744 cm$^{-1}$ is attributed to ester carbonyl stretching, indicating the presence of triglycerides and supporting the ability to detect and differentiate lipid classes in saliva. Protein-related signals are clearly observed in the amide I and III regions: a $\beta$-sheet conformation appears at 1696 cm$^{-1}$, and C=O stretching bands are present at 1674 and 1670 cm$^{-1}$. The amide III band at 1312 cm$^{-1}$ further confirms the detection of protein content. Nucleic acids are identified by the C–N stretching of cytosine, detected around 1374–1375 cm$^{-1}$. A consistent but currently unassigned absorption band at 1769 cm$^{-1}$ was also noted, suggesting the presence of unidentified molecular components and highlighting the need for further analysis. Together, these spectral features provide a detailed molecular fingerprint of saliva, enabling comprehensive biochemical characterization of this complex biological fluid.

\section{Conclusions}\label{sec:conc}
In this study, we address the problem of COVID-19 detection from ATR-FTIR spectra of non-invasive saliva using our proposed CNN-BiLSTM technique. CNN-BiLSTM is a hybrid deep learning architecture that represents the combination of CNNs' robustness in extracting features from complex data and LSTMs' ability to handle temporal sequences. The technique was compared against a CNN, Transformer architecture, and other machine learning models, such as RF, XGB and SVM.

Experiments with real data demonstrated that the CNN-BiLSTM architecture was the most effective in distinguishing between control and infected samples, consistently outperforming the other models in terms of accuracy, precision, F1 score, and sensitivity. Specifically, the CNN-BiLSTM achieved an average accuracy of 0.80, while the CNN, which shares the same configurations except for the BiLSTM layer and obtained the second best performance, achieved an average accuracy of 0.76.

The superiority of the CNN-BiLSTM highlights the importance of incorporating temporal information into modeling, which is crucial for analyzing ATR-FTIR spectra. Other machine learning models showed greater variation in results, indicating the need for additional preprocessing, hyper-parameter tuning or representation learning techniques to improve their stability and performance.

As potential future improvements, we aim to investigate attention mechanisms to improve the latent representation of our CNN-BiLSTM technique for analysis of ATR-FTIR spectra, and to explore alternative classification models or ensemble learning. Additionally, investigating advanced explainability techniques could provide more detailed insights about the biochemical features that influence COVID-19 detection.

In summary, our work demonstrates that combining ATR-FTIR spectroscopy with hybrid deep learning techniques, particularly the CNN-BiLSTM, represents a promising approach for medical diagnostics. This combination offers a rapid, portable, accurate, and sustainable (reagent-free) alternative for detecting infectious diseases, with potential applications in high-demand testing environments such as hospitals, airports and workplaces.

\bibliographystyle{splncs04}
\bibliography{refs}

\end{document}